\documentstyle[psfig,aps]{revtex}
\tightenlines
\def\Journal#1#2#3#4{{#1} {\bf #2}, #3 (#4)}

\def\AP{{\em Ann. Phys.}}
\def\CMP{{\em Commun. Math. Phys.}}

\def\HPA{{\em Helv. Phys. Act.}}
\def\IJMPB{{\em Int. J. Mod. Phys.} B}

\def\JSP{{\em J. Stat. Phys.}}

\def\PLB{{\em Phys. Lett.} B}
\def\PM{{\em Philos. Mag.}}

\def\PR{{\em Phys. Rev.}}
\def\PRA{{\em Phys. Rev.} A}
\def\PRB{{\em Phys. Rev.} B}
\def\PRD{{\em Phys. Rev.} D}

\def\RMP{{\em Rev. Mod. Phys.}}
\def\RMAP{{\em Rev. Math. Phys.}}

\newcommand{\be}{\begin{equation}}
\newcommand{\ee}{\end{equation}}
\newcommand{\bea}{\begin{eqnarray}}
\newcommand{\eea}{\end{eqnarray}}
\newcommand{\hf} {{1\over2}}
\newcommand{\nonu}{\nonumber\\}
\def\la{\langle}
\def\ra{\rangle}
\def\ord#1{{\cal O}(#1)}
\def\eq#1{(\ref{#1})}

\def\psid{\psi^\dagger}
\def\cD{{\cal D}}
\def\fd#1#2{{\delta{#1}\over\delta{#2}}}
\def\fdd#1#2#3{{\delta^2{#1}\over\delta{#2}\delta{#3}}}
\def\ke{{\cal E}}

\def\tG{\tilde\Gamma}
\def\jd{j^\dagger}
\def\grad{=^{\hskip -0.27cm\mathrm{gr}}}
\def\tn{{\text{nd}}}
\def\tf{{\text{fl}}}
\def\tr{{\text{Tr}}}

\begin{document}
\title{Functional Callan-Symanzik Equation for the Coulomb Gas}
\author{Sebastiao Correia$^a$\thanks{correia@lpt1.u-strasbg.fr},
Janos Polonyi$^{ab}$\thanks{polonyi@fresnel.u-strasbg.fr},
Jean Richert$^a$\thanks{richert@lpt1.u-strasbg.fr}}
\address{$^a$Laboratoire de Physique Th\'eorique\thanks{Unit\'e Mixte
de Recherche CNRS-Universit\'e, UMR 7085}, Universit\'e Louis Pasteur\\
3 rue de l'Universit\'e 67084 Strasbourg Cedex, France}
\address{$^b$Department of Atomic Physics, L. E\"otv\"os University\\
P\'azm\'any P. S\'et\'any 1/A 1117 Budapest, Hungary}
\date{\today}
\maketitle
\begin{abstract}
A non-perturbative scheme, based on the functional generalization
of the Callan-Symanzik equation is developed to treat 
the Coulomb interaction in an electron gas. The one-particle
irreducible vertex functions are shown to satisfy an evolution
equation whose initial condition is given by means of the classical 
action and the final point corresponds to the 
physical system. This equation is truncated by expanding
it in momenta and excitation energies,
leaving the electric charge as an arbitrary, not necesseraly small
parameter.
Exact coupled partial differential equations up to first order in the
frequencies and excitation energies are derived.  The numerical
integration of these eqations is left to a later stage.  Nevertheless,
in order to demonstrate the relation with the perturbation expansion
the one-loop Lindhard function and screening are reproduced
in the independent mode approximation of the evolution equation.
\end{abstract}

\section{Introduction}
In the many electron problem the Coulomb interaction is usually 
taken into account by a partial resummation of the perturbation 
expansion. A possible non-perturbative approach,
based on the concepts of the renormalization group and the
Callan-Symanzik equation is proposed in this paper. Originally viewed
as a physically motivated partial resummation of the perturbative
contributions \cite{earg}, the renormalization group strategy has
lately developed into a general method to treat strongly coupled systems
\cite{weha}-\cite{int}. The advantage of this method comes from
the functional treatment of the renormalization group
flow equation where the only approximation is the truncation
of the effective action according to the gradient expansion.
This formalism, originating from the Wegner-Houghton equation
\cite{weha}, allows us to follow a large number of coupling
constants and makes the approximation rather flexible. Although the
Callan-Symanzik \cite{casy} equation is {\em not} a renormalization group
equation because it traces the changes of the dynamics as
the mass is varied, its appearance and universality,
assumed that the mass is the only scale in the UV regime, 
show similarity with the renormalization group method.

The traditional application of the renormalization group method
for fermions with finite density \cite{bega}-\cite{salm} represents a
promising approach for the treatment of the Coulomb interaction
in an electron gas. We present in this paper a
version of the functional renormalization group method
for systems with Fermi surface which allows for
the systematic improvement of the scheme followed in these works.
The complication of the blocking in
fermionic systems comes from the fact that the successively
eliminated modes should be concentrated more and more at the
Fermi surface as we follow the renormalization group flow. In the
zero density case the infrared limit consists of a single homogeneous
mode. But the modes eliminated by the usual blocking
in the infrared regime of fermionic systems with finite density populate
the Fermi surface. This poses a problem for the implementation of the 
usual blocking procedure where the interactions are represented by 
fields. On the one hand, the momentum exchange between electrons 
is kept finite, of order of $k_F$ in the infrared limit, which requires the
evolution of these coupling strength. On the other hand, the scalar
fields are reduced to the point $p=0$ in the infrared limit and their
coupling strength at the scale $k_F$ is already frozen out. In other words,
the momentum modes of the field variables, responsible of the Coulomb
interaction can not be eliminated in a sequential step-by-step
manner during the blocking and one can not avoid non-local interactions.

In order to overcome this difficulty we do not order
the modes for their elimination according to their
energy. Instead, our blocking will eliminate the quantum fluctuations
in the increasing order of their amplitude \cite{int}. In this manner
we can keep the fundamental requirement of the blocking,
namely that it starts with the more weakly coupled modes and
treats the more strongly coupled ones later in the renormalization group
flow, without paying a high price for the presence of a Fermi surface.

The strategy of this generalized renormalization group method can
briefly be summarized as follows. First, one introduces a
parameter $M$ in the Hamiltonian in such a way that $H\to H_M$
where $H_M$ is almost diagonal in the free particle basis for large
$M$, i.e. the interactions are suppressed
as $M\to\infty$. The next step is the computation of the
evolution equation in $M$ for the effective action, the
generator functional of the one-particle irreducible (1PI) vertices.
The renormalization group improved approach consists of a
systematic expansion in the vicinity of the current system, namely
considering $H_M$ as the non-perturbed Hamiltonian and using
$\Delta H_M=H_{M-\Delta M}-H_M$ as perturbation
to obtain the changes in the effective action as $M$
is decreased by $\Delta M$. Notice that the exact
evolution equation can be obtained already in the leading order of
the perturbation expansion as $\Delta M\to0$ because
$\Delta H_M=\ord{\Delta M}$, in other words, $\Delta M/M$ is a new small
parameter to obtain a differential equation for the $M$-dependence.
Once the evolution equation is known, its
integration from the perturbative initial condition of the
system from large $M$ down to $M=0$ produces the desired
1PI vertices. The evolution equation, being a differential
equation for a functional, becomes useful only after projecting
it onto a restricted subspace of functionals. In this subspace
the effective action is characterized by ordinary functions and the
evolution equation is converted into a set of coupled differential
equations for them.

This method is based on the following two general assumptions:
(i) The dependence of the 1PI functions on the control parameter
is supposed to be differentiable. Notice that this condition,
the differentiable "turning on" the Coulomb interaction
is weaker than the assumption needed for the applicability
of the perturbation expansion, the analyticity for arbitrary values of
the control parameter. (ii) The possibility of using a simple
ansatz for the effective action to incorporate the physics of the model.
This step turns the evolution equation, a functional differential
equation into a set of coupled differential equations.
With these assumptions
the perturbation expansion around the Gaussian system is avoided,
the 1PI functions are obtained by solving differential equations
rather than computing Feynman graphs.

The parameter $M$ controls the amplitude of the quantum
fluctuations. In a relativistic system the mass plays such a role
and the simplest realization of the blocking in the amplitude
is based on the modification $m^2\to m^2+M^2$ of the
Hamiltonian. The resulting differential equation is the
functional generalization of the Callan-Symanzik equations\cite{casy}.

In this paper we present the first step only in the implementation
of these ideas in the framework of the Coulomb gas. We consider
1PI functions with the following restrictions:
(i) The dependence on the external energies and momenta
is kept up to $\ord{\omega}$ and $\ord{\ke}$ for electrons where $\omega$ is the frequency and $\ke$
is the energy counted from the Fermi surface and up to $\ord{\omega^2}$
and $\ord{p^2}$ for photons.
(ii) There are no more than two external electron legs. The number
of photons is not restricted.

We arrive at a set of coupled differential equations in this manner
for the functions parametrising the 1PI functions. The numerical
integration in the control parameter yields the solution of the
Coulomb problem for the assumed ansatz for the 1PI functions.
There is no problem in principle to generalize the scheme and
include 1PI functions with more external electron lines and
more complicated energy and momentum dependence.

In order to demonstrate the relation between the perturbation
expansion and our scheme we integrate out the evolution equation
in the independent mode approximation. In this approximation
the system remains unchanged during the evolution, meaning
that the modes are taken into account independently from each
other. This corresponds to the one-loop approximation of the
perturbation expansion. The one-loop Thomas-Fermi screening
and Lindhard function are reproduced in our scheme without
considering Feynman graphs.

We derive the evolution equation for the electron gas with Coulomb
interaction in section II. The resulting
functional differential equation is projected on the leading order
gradient expansion ansatz (expansion in $\omega$, $\ke$ and $p$) for the effective action in section III.
In order to check the consistency of this scheme with the
standard perturbation expansion we solve the evolution equations
analytically in the independent mode approximation in section IV.
Section V contains the conclusions. Some technical details of the
functional derivation and the gradient expansion are shown in the
Appendices.

\section{Evolution equation}
We shall study spinless electrons interacting by means of the
Coulomb potential and described by the Hamiltonian
\be
H=\int d^3x\psid(x)\left[-{\hbar^2\over2m}\Delta-\mu\right]\psi(x)
+\int d^3xd^3y\psid(x)\psid(y){e^2\over4\pi|x-y|}\psi(y)\psi(x).
\ee
where the field $\psi$ corresponds to electrons and $\mu$ is the
chemical potential.
In the path integral formalism we introduce an auxiliary field
$u(x,t)$, taken formally as the temporal component of the photon field,
write the action  in terms of the Grassmannian $\psi$, $\psid$
and real $u$ as
\be\label{bare}
S[\psi,\psid,u]=\int_x\left\{\psid_x\left[i\hbar\partial_t+
{\hbar^2\over2m}\Delta+\mu+eu_x\right]\psi_x
-\hf(\nabla u_x)^2\right\}
\ee
in the functional integral
\be
\int\cD[\psid]\cD[\psi]\cD[u]e^{{i\over\hbar}S[\psi,\psid,u]}.
\ee
The $x,t$ or $p,\omega$ arguments of the
field variables will be denoted by the subscripts $x$ and $p$,
respectively. The symbols $x$ and $p$ stand for the spatial
components only if written otherwise. In a similar manner,
we shall use the notation $\int_x=\int dtd^3x=VT$,
$\int_p=\int d\omega d^3p/(2\pi)^4$, and
$f_x=\int_qe^{iq\cdot x-i\omega t}f_q$.
The field $u_x$ can be considered either as an auxiliary field,
responsible for the Coulomb interaction or the temporal
component of the photon field in the Feynman gauge after both
the transverse and longitudinal components of the
spatial photon field have been suppressed.
The homogeneous mode of the photon field,
\be\label{unul}
u_0={1\over VT}\int_xu_x
\ee
plays a special role because of the absence of $\ord{u_0^2}$ terms
in the action. In fact, the homogeneous component of the photon
field $u_0$ contributes to the chemical potential modifying
the electron density and the inhomogeneous field components
induce polarization effects only. Another way to see this is to note
that the integration over the $u_0$ configuration
induces the constraint $\int_x\psid_x\psi_x=0$
and wipes out asymptotic states with non-vanishing electric charge.
Actually, the homogeneous photon field configurations in QED,
$A_\mu(x,t)=a_\mu$ are not to be integrated over
in the path integral of a non-confining gauge theory \cite{kreutz}.
Thus the integral measure $\cD[u]$ does not
involve the integration over the homogeneous mode i.e. $u_0=0$.

We add a quadratic term to the action, $S\to S_M=S+S_s$,
\be\label{suppr}
S_s[\psi,\psid,u]=\int_x\left\{f(M^2)\psid_x
\left({\hbar^2\over2m}\Delta+\mu\right)\psi_x
-{M^2\over2}u^2_x\right\},
\ee
in order to control the quantum fluctuations.
The function $f(M^2)$ is chosen in such a manner that $f(0)=0$ and
$\lim_{M\to\infty}f(M^2)=\infty$, in order
to  suppress the fluctuations of the fields when $M\to\infty$.
In practice we shall use the simple form $f(M^2)=M^2/m^2$.
Since we are interested in a system of particles at finite
density, the chemical potential $\mu$ is non-vanishing. This
particular form, \eq{suppr}, was chosen to control the
strength of interactions by changing the weight of the
quadratic, non-interacting part of the action. The gradual
decrease of $S_s$ "turns on" the interactions and the
integration of the evolution equation in the control parameter
allows us to solve the interactive system. The suppression action
can be chosen to be gauge invariant, but we are satisfied with
this simpler form because the restriction for the Coulomb interaction
only, the suppression of the spatial component of the photon field
alone removes the gauge symmetry.

The action $S_M$ determines the dependence of the
microscopic physics on the control parameter $M$. In order to trace
down the role played by $M$ we consider the generator functional of the
1PI vertices, the effective potential defined by
\be\label{legtr}
\tG_M[\psi,\psid,u]=W_M[j,\jd,J]-\jd\cdot\psi
-\psid\cdot j-J\cdot u
\ee
where $f\cdot g=\int_xf_xg_x$,
\be\label{conngf}
\tilde Z_M[j,\jd,J]=e^{{i\over\hbar}W_M[j,\jd,J]}
=\int\cD[\psid]\cD[\psi]\cD[u]e^{{i\over\hbar}
\left(S_M[\psi,\psid,u]+\jd\cdot\psi+\psid\cdot j
+J\cdot u \right)},
\ee
and
\bea\label{vartr}
\fd{W_M[j,\jd,J]}{j_x}
&=-\la\psid_x\ra_{j,\jd,J}&=-\psid_x,\nonu
\fd{W_M[j,\jd,J]}{\jd_x}
&=\la\psi_x\ra_{j,\jd,J}&=\psi_x,\nonu
\fd{W_M[j,\jd,J]}{J_x}
&=\la u_x\ra_{j,\jd,J}&=u_x.
\eea
We use the same notation for the quantum field and its
expectation value.

We first derive a functional differential equation in $M^2$ for the
effective action,
\be\label{eveq}
{d\tG_M\over d(M^2)}={\cal F}[\tG_M;M]
\ee
whose integral,
\be\label{solev}
\tG_0=\tG_M-\int_0^{M^2}d(M')^2{\cal F}[\tG_{M'};M'],
\ee
yields the effective action of the model we want to solve in terms
of the initial condition
$\tG_M[\psi,\psid,u]= S_M[\psi,\psid,u]$
imposed when the $M^2$ exceeds every energy scale of the model.

The evolution equation of the effective action can be written
\cite{int} as
\be\label{evgw}
\partial_{M^2}\tG_M[\psi,\psid,u]=\partial_{M^2}W[j,\jd,J],
\ee
where $\partial_{M^2}=\partial/\partial M^2$ according to \eq{legtr}
and \eq{vartr}. Since we know the control
parameter dependence of the bare action we have immediately
\bea\label{eveqw}
\partial_{M^2}W_M&=&-i\hbar e^{-{i\over\hbar}W_M}
\partial_{M^2}e^{{i\over\hbar}W_M}\nonu
&=&-f'(M^2)\tr\la\psid_x\ke\psi_x\ra_{\jd,j,J}
-\hf \tr\la u_x^2\ra_{\jd,j,J},
\eea
where $f'=\partial_{M^2}f$, $\ke=-\hbar^2\Delta/2m-\mu$
measures the kinetic energy operator measured from the Fermi level
and $\tr$ denotes the functional trace, $\tr A=\int_xA_{x,x}$.

We introduce now the notation
\be
F^{(1)}_f=\fd{F[f]}{f},~~~F^{(2)}_{fg}=\fdd{F[f,g]}{f}{g},
\ee
to relate the second functional derivatives of $W$ and $\tG$
\be
\pmatrix{\fd{\psid_z}{\psid_x}&\fd{\psi_z}{\psid_x}&\fd{u_z}{\psid_x}\cr
\fd{\psid_z}{\psi_x}&\fd{\psi_z}{\psi_x}&\fd{u_z}{\psi_x}\cr
\fd{\psid_z}{u_x}&\fd{\psi_z}{u_x}&\fd{u_z}{u_x}}
=\int_y
\pmatrix{\tG^{(2)}_{\psid_x\psi_y}
&\tG^{(2)}_{\psid_x\psid_y}&\tG^{(2)}_{\psid_xu_y}\cr
\tG^{(2)}_{\psi_x\psi_y}&\tG^{(2)}_{\psi_x\psid_y}&\tG^{(2)}_{\psi_xu_y}\cr
\tG^{(2)}_{u_x\psi_y}&\tG^{(2)}_{u_x\psid_y}&\tG^{(2)}_{u_xu_y}}
\pmatrix{-W^{(2)}_{\jd_yj_z}
&W^{(2)}_{\jd_y\jd_z}&W^{(2)}_{\jd_yJ_z}\cr
W^{(2)}_{j_yj_z}&-W^{(2)}_{j_y\jd_z}&-W^{(2)}_{j_yJ_z}\cr
W^{(2)}_{J_yj_z}&-W^{(2)}_{J_y\jd_z}&-W^{(2)}_{J_yJ_z}}
=\delta_{x,z}\cdot\mathbf{1}
\ee
The evolution equation \eq{evgw} goes over to
\be
\partial_{M^2}\tG_M[\psid,\psi,u]
=-i\hbar f'(M^2)\tr[\ke W^{(2)}_{j,\jd}]-f'(M^2)\tr[\ke\psid\psi]
+{i\hbar\over2}\tr[W^{(2)}_{J,J}]-\hf \tr[u^2].
\ee
In order to remove the bilinear terms in the fields on the right hand side
we write out explicitly the artificial suppression terms in the effective action,
\be
\tG_M[\psid,\psi,u]=\Gamma_M[\psid,\psi,u]
-\int_x\left(f(M^2)\psid_x\ke\psi_x+\hf M^2 u^2_x\right),
\ee
which leads to
\be\label{eveqg}
\partial_{M^2}\Gamma_M[\psid,\psi,u]=-i\hbar f'(M^2)\tr
\left[\ke\left(\Gamma^{(2)}_M-{\cal M}^2
\right)^{-1}_{j,\jd}\right]
+{i\hbar\over2}\tr\left[\left(\Gamma^{(2)}_M-{\cal M}^2
\right)^{-1}_{J,J}\right],
\ee
a closed evolution equation for the effective action, with
\be
{\cal M}^2=\pmatrix{-\ke&0&0\cr0&\ke&0\cr0&0&M^2\cr}.
\ee

Two remarks are in order about details not shown explicitly in
Eq. \eq{eveqg}:

\begin{itemize}
\item One should bear in mind that the right hand side
is UV divergent as it stands because the suppression of the
amplitude of the fluctuations is not a regulator. In order to
have well defined expressions we introduce a cut-off
$\Lambda$ for the momentum integrals.
This regulator will be kept finite because our non-relativistic
model is non-renormalizable.

\item We are working in real time, and the quadratic part of the action
must be supplied with an infinitesimal imaginary part. We shall use the
electron propagator
\be\label{elprop}
G(p,\omega)={e^{i\omega\eta'}\over\omega-{p^2\over2m}
+\mu+i\delta_p}=\left\{{\Theta(k_F-p)\over\omega-
{p^2\over2m}+\mu-i\delta}+{\Theta(p-k_F)\over\omega-{p^2\over2m}
+\mu+i\delta}\right\}e^{i\omega\eta'},
\ee
where $\eta'\to0^+$.
Since the chemical potential dependence is not always continuous
we have to make sure that $k_F$ does not evolve with the control
parameter $M^2$ in order to avoid the appearance of singularities in the
evolution equation. We shall achieve the $M$-independence of the
Fermi-momentum by the choice $\delta_p=0^+\text{sign}(p-k_F)$.
In the case when the chemical potential dependence is continuous then
schemes with an $M$-dependent Fermi-momentum are possible but
their evolution equation appears with an additional term,
$\partial\mu/\partial M$ times the density in \eq{eveqw}.
\end{itemize}

\section{Gradient expansion}
\subsection{Effective action}\label{sec:effective-action}
In order to convert the functional differential equation \eq{eveqg}
into a set of coupled differential equation for the running
coupling constants which are the parameters of the effective action
we introduce an ansatz for $\Gamma_M[\psid,\psi,u]$.
We first separate the fluctuations of the photon field from the
homogeneous background,
\be\label{phf}
u_x=u_0+\eta_x,
\ee
where $\eta_{p=0}=0$. Note that once an ansatz is introduced
for the effective action we may explore its structure for
$u_0\not=0$ even if homogeneous photon field configurations are not
allowed in the path integral. Thus $u_0\not=0$ will be used
in the effective action whose form will be written as
\bea\label{grexf}
\Gamma_M[\psid,\psi,u]&=&\int_x\psid_x\biggl[
iX_0(u_x)\partial_t-X_1(u_x)\ke+U_1(u_x)+i\delta_\ke
+\hf Y_0(u_x)(\partial_tu_x)^2+\hf Y_1(u_x)(\nabla u_x)^2\biggr]\psi_x\nonu
&&+\int_x\left[\hf Z_0(u_x)(\partial_tu_x)^2-\hf Z_1(u_x)(\nabla u_x)^2-U_2(u_x)
+i\epsilon u_x^2\right],
\eea
where $\delta_\ke=0^+\mathrm{sign}\ke$ and the $X_i$, $Y_i$, 
$Z_i$ and $U_i$ are polynomials in the field $u_x$.
This funtional form is given in the framework of the gradient
expansion which is an expansion in the
space-time derivatives of the photon field truncated
at $\ord{\partial^2}$. We write the effective action in terms
of the 1PI functions in momentum space,
\bea\label{genfunc}
\Gamma[\psid,\psi,u_0]&=&\int_{p_1,p_2}\Gamma^{\psid\psi}_{p_1,p_2}
\psid_{p_1}\psi_{p_2}
+\int_{p_1,p_2,q}\Gamma^{\psid\psi u}_{p_1,p_2,q}\psid_{p_1}\psi_{p_2}u_q\nonu
&&+\hf\int_{q_1,q_2}\Gamma^{uu}_{q_1,q_2}u_{q_1}u_{q_2}
+\hf\int_{p_1,p_2,q_1,q_2}\Gamma^{\psid\psi uu}_{p_1,p_2,q_1,q_2}
\psid_{p_1}\psi_{p_2}u_{q_1}u_{q_2}+\cdots
\eea
where the superscripts show the type of the external legs
and the subscripts give the corresponding momenta. One finds by using 
this form that the gradient expansion corresponds to the Taylor expansion
of the 1PI functions in the photon energy-momentum. In the present
work the electronic 1PI functions will be considered by expanding in
the frequency and the kinetic energy, $\partial_t$ and $\ke$,
respectively.
The dependence on the Grassmannian fermionic field variables is kept
quadratic in the present approximation which reduces the argument of
the local coefficient functions $X$, $Y$, $Z$ and $U$
to the photon field $u$.

The potentials $U_1(u)$ and $U_2(u)$ are space-time independent,
indicating that they do shift the contour of the frequency integrals.
The finite lifetime effects might come from the imaginary part of the
electron self energy. This appears in $\ord{\omega^2}$ and will be
ignored in the present work which includes $\ord{\omega}$ only.

The infrared divergences of the perturbation
expansion generate singular behaviour at $\omega=p=0$. This fact
introduces some complications. In particular, the photon 1PI
functions are not continuous at this point
and their limiting value depends on the way this point is approached,
as the remnant of gauge invariance, the transversality of the photon
self energy.
This feature introduces an important difference between the
$\ord{\eta^0}$ and the $\ord{\eta^2}$ level solution of the
evolution equations. We have already mentioned after Eq. \eq{unul}
that the dynamics of the homogeneous
mode of the photon field is different from the non-homogeneous
ones. To understand better the way this appears in our scheme one must
remember that the solution of the complete functional evolution equation
for the effective action generates the exact 1PI functions, cf.
Eq. \eq{genfunc}. As we shall show below, the
truncated gradient expansion ansatz for the effective action
reduces the evolution equation into a hierarchy of approximate relations
between the 1PI functions because only a polynomial of finite order
in the external momenta is retained for the 1PI functions in the
otherwise exact equation. The integration of the contributions of order
$\ord{\eta^0}$ to the evolution equation reproduces the value
of the 1PI functions at $\omega=p=0$ in this approximation.
For example, the integration of the terms $\ord{\eta^2}$ yield
the photonic two point 1PI function at arbitrary
energy-momenta $\Gamma^{appr}_{uu}(\omega_1,p_1,\omega_2,p_2)$, but
\be\label{disc}
\lim_{\omega,p\to0}\Gamma^{appr}_{uu}(\omega,p,-\omega,-p)
\not=\Gamma^{appr}_{uu}(0,0,0,0).
\ee
We take such a non-continuous energy-momentum dependence of the
effective action into account by replacing the potential $U_2(u_x)$
in \eq{grexf} by two different functions,
\be\label{u2bar-u2}
\int_xU_2(u_x)\to\cases{\int_xU_2(u_0+\eta_x)&$\eta\not=0$,\cr
VT\bar U_2(u_0)&$\eta=0$.}
\ee
The first line on the right hand side, representing the
left hand side in Eq. \eq{disc}, will be evaluated by
taking the limit $\omega\to0$ before $p\to0$, cf. Eq. \eq{lijon}.\\

For large enough $M$, when the quantum fluctuations are negligible,
we have the initial condition
\be\label{initc}
X_0=\hbar,~~X_1=Z_1=1,~~Y_0=Y_1=U_2=\bar U_2=0,~~U_1=eu,
\ee
up to $\ord{M^{-2}}$ corrections because $\Gamma_M[\psid,\psi,u]=S[\psid,\psi,u]$, see after Eq.~(\ref{solev}).

\subsection{Expansion in the photon field fluctuations}
In order to determine the evolution equations for the coefficient
functions we write the photon field as in Eq. \eq{phf}
and expand Eq. \eq{eveqg} in powers of $\eta_x$. The left hand side is
\be
\Gamma_M[\psid,\psi,u]=\Gamma_M[\psid,\psi,u_0]+\int_z\eta_z
\Gamma^{(1)}_{Mu_z}[\psid,\psi,u_0]
+\hf\int_{z,z'}\eta_z\eta_{z'}
\Gamma^{(2)}_{Mu_z u_{z'}}[\psid,\psi,u_0]+\ord{\eta^3},
\ee
when $\eta\not=0$. The zeroth order term in $\eta$ gives
\be\label{eq:gamma-u0}
\Gamma_M[\psid,\psi,u_0]=\int_q\psid_{-q}\psi_qA_q-VT\bar U_2,
\ee
where $u_0$ is an arbitrary parameter, introduced in order to
explore our functional differential equation \eq{eveqg},
\be\label{eq:Aq}
A_q=X_0\omega_q-X_1\ke_q+U_1,
\ee
the coefficient functions are taken at $u_0$, and $\ke_q=q^2/2m-\mu$.
Note that even though we have $u=u_0$ in Eq. \eq{eq:gamma-u0}
we are dealing with the $\eta\not=0$ part of the effective action
because the functional Taylor expansion coefficients are
obtained in the limit $\eta\to0$.
The contributions to the first and second order in $\eta$ can be
written as
\be
\int_{q_1,q_2,q_3}\eta_{q_1}\psid_{q_2}\psi_{q_3}
A_{q_3}^{(1)}\delta_{q_1+q_2+q_3,0}
\ee
and
\be
\int_{q_1,q_2,q_3,q_4}\eta_{q_1}\eta_{q_2}\psid_{q_3}\psi_{q_4}
\left[A_{q_4}^{(2)}-Y_0\omega_{q_1}\omega_{q_2}-Y_1q_1\cdot q_2\right]
\delta_{q_1+q_2+q_3+q_4,0}
-\int_{q}\eta_q\eta_{-q}\left[Y_0\omega_q^2+Y_1q^2+U^{(2)}_2\right]
\ee
by means of the functional derivatives \eq{firstfd} and \eq{secondfd}
recorded in the Appendix.

On the right hand side of the evolution equation~(\ref{eveqg}) we find the second
functional derivative matrix
$\tG^{(2)}_M=\tG_{\text{d}}+\Gamma_{\tn}+\Gamma_{\tf}$,
where $\tG_{\text{d}p_1,p_2}$ is vanishing for $p_1+p_2\not=0$,
$\Gamma_{\tn}=\ord{\eta^0}$ and $\Gamma_{\tf}$
comprises the $\ord{\eta}$ and $\ord{\eta^2}$ contributions.
The $\ord{\eta^0}$, diagonal piece is obviously
\be
\tG_{\text{d}}=\pmatrix{-G^{-1}_{p_2}&0&0\cr0&G^{-1}_{p_1}&0\cr
0&0&D^{-1}_{p_1}}\delta_{p_1+p_2,0},
\ee
where
\bea\label{fbprop}
G^{-1}_p&=&A_p-f(M^2)\ke_p+i\delta_{\ke_p},\nonu
D^{-1}_p&=&Z_0\omega^2_p-Z_1p^2-U_2^{(2)}-M^2+i\epsilon.
\eea
Note that the $U_1$ of Eq. \eq{eq:Aq} does not
influence the imaginary part of $G$ and the Fermi level remains
independent of the photon field. This is in agreement with our
introduction of $u_0$ in the effective action as a formal parameter
to help to uncover the content of the functional equation \eq{eveqw}
rather than a dynamical variable.

The remaining $\ord{\eta^0}$ contributions are easily found,
\be
\Gamma_\tn=\pmatrix{0&0&\Sigma^{\psid u}\cr0&0&\Sigma^{\psi u}\cr
\Sigma^{u\psi}&\Sigma^{u\psid}&\Sigma^{uu}}
\ee
the matrix elements being listed in Eq. \eq{abcde}. The terms
$\ord{\eta}$ and $\ord{\eta^2}$, parametrized as
\be\label{parsecse}
\Gamma_{\tf}=\pmatrix{\sigma^{\psid\psi}&\sigma^{\psid\psid}&\sigma^{\psid u}\cr
\sigma^{\psi\psi}&\sigma^{\psi\psid}&\sigma^{\psi u}\cr
\sigma^{u\psi}&\sigma^{u\psid}&\sigma^{uu}}+\ord{\eta^3},
\ee
are explicitly presented in Eq. \eq{fluctsee}.

The inversion of $\Gamma^{(2)}$ is carried out by expanding in
the non-diagonal pieces,
\bea\label{neum}
W^{(2)}&=&W^{(2)}_0
+\sum_{\alpha_1}W^{(2)}_{\alpha_1}
+\sum_{\alpha_1,\alpha_2}W^{(2)}_{\alpha_1,\alpha_2}
+\sum_{\alpha_1,\alpha_2,\alpha_3}W^{(2)}_{\alpha_1,\alpha_2,\alpha_3}
+\sum_{\alpha_1,\alpha_2,\alpha_3,\alpha_4}
W^{(2)}_{\alpha_1,\alpha_2,\alpha_3,\alpha_4}+\cdots
\eea
where the summation is over the indices $\alpha_j=\tn$ or $\tf$ and
$W^{(2)}_0=\Gamma_{\text{d}}^{-1}$ denotes the diagonal part of the
propagator. The non-diagonal contributions are collected in
\bea
W^{(2)}_{\alpha_1}&=&-W^{(2)}_0\Gamma_{\alpha_1}W^{(2)}_0,\nonu
W^{(2)}_{\alpha_1,\alpha_2}&=&W^{(2)}_0\Gamma_{\alpha_1}
W^{(2)}_0\Gamma_{\alpha_2}W^{(2)}_0,\nonu
W^{(2)}_{\alpha_1,\alpha_2,\alpha_3}&=&W^{(2)}_0\Gamma_{\alpha_1}
W^{(2)}_0\Gamma_{\alpha_2}W^{(2)}_0\Gamma_{\alpha_3}W^{(2)}_0,\nonu
W^{(2)}_{\alpha_1,\alpha_2,\alpha_3,\alpha_4}&=&-W^{(2)}_0\Gamma_{\alpha_1}
W^{(2)}_0\Gamma_{\alpha_2}W^{(2)}_0\Gamma_{\alpha_3}
W^{(2)}_0\Gamma_{\alpha_4}W^{(2)}_0.
\eea

In the last step we substitute the gradient expanded form, \eq{grexf},
and identify in \eq{eveqg} the contributions of the different powers in
the electron field and the fluctuations in order to get the evolution
equations for the coefficient functions.

\subsection{Homogeneous photon field}
The zeroth order in the fluctuations is obtained by setting
$\eta=0$ on both sides of \eq{eveqg}. On the right hand side,
in particular in Eq. \eq{neum} this amounts to neglect
terms containing $\Gamma_{\tf}$. The rest involves the expressions
\bea
\tr
\left[\ke\left(\Gamma^{(2)}_M-{\cal M}^2
\right)^{-1}_{j,\jd}\right]=-\tr[\ke W^{(2)}_{j,\jd}]&=&\int_p\ke_pG_p
-\int_p\int_{p'}\ke_pG_p^2\Sigma^{\psi,u}_{-p,p'}D_{p'}
\Sigma^{u,\psid}_{-p',p},
\eea
and
\bea
\tr\left[\left(\Gamma^{(2)}_M-{\cal M}^2
\right)^{-1}_{J,J}\right]&=&\tr[W^{(2)}_{J,J}]\\
&=&\int_p\left[D_p-D_p^2\Sigma^{u,u}_{-p,p}\right]
+\int_{p,p'}D_p^2\biggl[\Sigma^{u,\psi}_{-p,p'}G_{p'}
\Sigma^{\psid,u}_{-p',p}
-\Sigma^{u,\psid}_{-p,p'}G_{p'}\Sigma^{\psi,u}_{-p',p}
-\Sigma^{u,u}_{-p,p'}D_{p'}\Sigma^{u,u}_{-p',p}\biggr].\nonumber
\eea
The left hand side of the evolution equation \eq{eveqg} includes
the expression \eq{eq:gamma-u0} and the use of \eq{fbprop} and
\eq{abcde} gives the evolution equations
\be\label{ukettoe}
\partial_{M^2}\bar U_2=i\hbar f'\int_p{\ke_p\over X_0\omega_p-(X_1+f)\ke_p+U_1}
-i\hbar\int_p{1\over Z_0\omega^2_p-Z_1p^2-U_2^{(2)}-M^2},
\ee
\bea\label{elevh}
&&\partial_{M^2}[X_0\omega_q-X_1\ke_q+U_1]\grad
-{i\hbar\over2}\int_p{A^{(2)}_q
+Y_0\omega_p^2+Y_1p^2\over(Z_0\omega_p^2-Z_1p^2-U_2^{(2)}-M^2)^2}\nonu
&&~~+{i\hbar\over2}A^{(1)}_q
\int_p{A^{(1)}_{-p}\over(Z_0\omega^2_{p+q}-Z_1(p+q)^2-U_2^{(2)}-M^2)^2
(X_0\omega_p-(X_1+f)\ke_p+U_1)}\nonu
&&+i\hbar f'A^{(1)}_q\int_p{\ke_pA^{(1)}_{-p}\over(X_0\omega_p
-(X_1+f)\ke_p+U_1)^2(Z_0\omega_{p+q}^2-Z_1(p+q)^2-U_2^{(2)}-M^2)}.
\eea
We introduced the symbol $\grad$
to indicate that the right hand side is supposed to be
written by expanding in the variables $\omega$, $\ke$ and $p$
and by retaining only the orders shown on the left hand side.

As we shall see below, the discontinuity discussed in Eq.~(\ref{disc})
affects only Eq.~(\ref{ukettoe}).  Eq.~(\ref{elevh}) gives the
evolution of the coefficents $X_0$, $X_1$ and $U_1$.  In order to
solve these coupled differential equations one needs other equations
for the $Z_i$'s, $Y_i$'s and $U_2$.

\subsection{Fluctuations}
In order to find the evolution of the coefficient functions
we substitute the gradient expanded form, \eq{grexf},
and identify on the right hand side of Eq. \eq{eveqg} the contributions
of different powers in $\psid\psi$ and $\eta$. Some expressions
of the contributions of the individual terms on the right hand side
of Eq. \eq{neum} are shown in Appendix \ref{neumap}.

It is easy to isolate the terms $\ord{(\psid\psi)^0\eta^2}$,
\bea\label{bnfeve}
&&\hf\left(\partial_{M^2}Z_0\omega^2_q-\partial_{M^2}Z_1q^2
-\partial_{M^2}U_2^{(2)}\right)\grad\nonu
&&~~~i\hbar f'\int_p{\ke_pA^{(1)}_pA^{(1)}_{p+q}\over(X_0\omega_p
-(X_1+f)\ke_p+U_1)^2(X_0\omega_{p+q}-(X_1+f)\ke_{p+q}+U_1)}\nonu
&&~~~+i{\hbar f'\over2}\int_p{\ke_p(A^{(2)}_p-Y_0\omega_q^2+Y_1q^2)
\over(X_0\omega_p-(X_1+f)\ke_p+U_1)^2}
-{i\hbar\over4}\int_p{Z_0^{(2)}(\omega_q^2+\omega_p^2)+Z_1^{(2)}
(q^2+p^2)+U_2^{(4)}\over(Z_0\omega^2_p-Z_1p^2-U_2^{(2)}-M^2)^2}\nonu
&&~~~-{i\hbar\over2}\int_p{[Z_0^{(1)}(\omega^2_p+\omega^2_q
+\omega_p\omega_q+Z_1^{(1)}(p^2+q^2+p\cdot q)+U^{(3)}_2]^2\over
(Z_0\omega^2_p-Z_1p^2-U_2^{(2)}-M^2)^2
(Z_0\omega^2_{p+q}-Z_1(p+q)^2-U_2^{(2)}-M^2)}.
\eea
The discontinuity \eq{disc} appears here as the non-commutivity of the
integration and the limit $\omega_q,q\to0$ in this equation.
The singularity at $q=0$ is caused by the fact that the order of
the singularity in $\omega_p$ is 2 or 3 when $q=0$ or $q\not=0$. Since
we discuss the $\ord{\eta^2}$ contributions we have to keep
$q\not=0$. The $q$-independent part of the right hand
side, the extrapolation from $q\not=0$ to $q=0$ will contribute
to $U_2^{(2)}$ rather than $\bar U_2^{(2)}$  according to
Eq.(\ref{u2bar-u2}).\\

The terms $\ord{\psid\psi\eta}$ on the left hand side of 
Eq.~(\ref{eveqg}) is
\be
\partial_{M^2}\int_{q_1,q_2}\eta_{q_1}\psid_{-q_1-q_2}\psi_{q_2}
\left[X_0^{(1)}\omega_{q_2}-X_1^{(1)}\ke_{q_2}+U_1^{(1)}\right].
\ee
Since we have $q_1\not=0$ there will be no singular coincidence of
the poles on the right hand side of the evolution equation. As a
result the terms $\ord{\omega_{q_2}^0\ke^0_{q_2}}$, $\ord{\omega_{q_2}\ke^0_{q_2}}$
and $\ord{\omega^0_{q_2}\ke_{q_2}}$ on the right hand side of the evolution
equation of order $\ord{\psid\psi\eta}$ reproduce the first
derivative of the right hand side of Eq. \eq{elevh} with respect to $u_0$.\\

In the order $\ord{\psid\psi\eta^2}$ we find
\be\label{etat} 
\hf\partial_{M^2}\int_{q_1,q_2,q_3}\eta_{q_1}\eta_{q_2} 
\psid_{-q_1-q_2-q_3}\psi_{q_3}\left[ 
X_0^{(2)}\omega_{q_3}-X_1^{(2)}\ke_{q_3}+U_1^{(2)}-Y_0\omega_{q_1}\omega_{q_2} 
-Y_1q_1\cdot q_2\right] 
\ee 
on the left hand side of the evolution equation.
The terms $\ord{\omega_{q_3}}$, $\ord{\ke_{q_3}}$ and $\ord{1}$
give the second derivatives of Eq. \eq{elevh} and can be omitted on
both sides of the equation. The remaining
contribution in $\ord{\omega_{q_1}\omega_{q_2}}$ and
$\ord{q_1\cdot q_2}$, obtained for $q_3=0$ is
\bea\label{hosszu}
&&\hf\partial_{M^2}\int_{q_1,q_2,q_3}\eta_{q_1}\eta_{q_2} 
\psid_{-q_1-q_2-q_3}\psi_{q_3}\left[Y_0\omega_{q_1}\omega_{q_2} 
+Y_1q_1\cdot q_2\right]\nonu
&&\grad i\hbar\left(\tr\left[\ke W_{\tf,j,\jd}\right]
-{\cal F}_2-{\cal F}_3-{\cal F}_4\right)
-f(M^2)\left(\tr\left[W_{\tf,J,J}\right]
+{\cal B}_2+{\cal B}_3+{\cal B}_4\right),
\eea
where the quantities on the right hand side are given in Appendix
\ref{neumap}. When the coefficients of the field variables with the
same arguments are identified on both side we find two differential
equations, one for $Y_0$ and another for $Y_1$. 

The form of these and the other differential equations for the 
coefficient functions can simply be understood in the following manner.
We start with the one-loop expressions for the 1PI vertex functions
using the non-Gaussian parts of the effective action $\tilde\Gamma$
as interaction. The infinitesimal
increase of the parameter $M^2\to M^2+\delta M^2$ modifies the
electron and photon propagators as
\be
G\to G-\delta M^2f'(M^2)G^2,\ \ \ D\to D-\delta M^2D^2,
\ee
respectively. The change induced by this transformation in the one-loop
expressions for the 1PI vertex functions
is the right hand side of our evolution equation up to the factor
$\delta M^2$. At this stage we recover the perturbation expansion
based on the Feynman rules. The usual accumulation effect of
the renormalization group strategy occurs when the differential
equations are integrated out. By making $n$ steps $M^2\to M^2+\delta M^2$
the integration sums up the $n$-loop contributions to
the corresponding 1PI vertex function.

In this manner the integration of the evolution equations Eqs. \eq{ukettoe},
\eq{elevh}, \eq{bnfeve} and \eq{hosszu} with the initial condition
\eq{initc} provides the effective action, the 1PI functions
in the given approximation of their dependence on the
external momenta. The successive inclusion of terms $\ord{(\psid\psi)^n}$
or higher order derivatives in the ansatz \eq{grexf} for the effective action
yields more complete solutions.

\section{Independent mode approximation}
The renormalization group strategy is based on the accumulation of the
effects of the modes which are eliminated sequentially. This corresponds to the
integration of the evolution equations where the right hand side
evolves with $M^2$. We can easily
obtain an analytic approximation for the evolution by ignoring the
$M^2$-dependence on the right hand side of the evolution equation.
This amounts to the
independent mode approximation which reproduces the one-loop 1PI
functions in the given order of the external momenta.

\subsection{Homogeneous photon field: Hartree-Fock approximation}
As already mentioned in paragraph~\ref{sec:effective-action}, the 1PI
functions are non-analytic at $\omega=p=0$.  We shall first show
that this implies that $\bar U_2$ does not evolve. The right hand side of Eq.
\eq{eveqg} is obtained by using an $M^2$-independent effective action \eq{grexf},
given by Eq. \eq{initc}. As far as the $\eta=0$ case is concerned we
find
\be\label{imhb}
\partial_{M^2}\bar U_2=-i\hbar f'\int_p{\ke_p\over\hbar\omega_p
-(1+f)\ke_p+eu_0}-i\hbar\int_p{1\over p^2+M^2},
\ee
After the frequency integration of the first term,
\be
\lim_{\eta\to0}\int{d^3p\over(2\pi)^3}\ke_p\int{d\omega\over2\pi}
{e^{i\omega\eta}\over\hbar\omega_p-(1+f)\ke_p+eu_0+i\delta_p}
=\int{d^3p\over(2\pi)^3}\ke_pi\Theta(-\ke_p) =-i{k_F^5\over30\pi^2m},
\ee
the right hand side turns out to be $u_0$-independent. Since
we are interested in the field dependent part of the effective action
only, Eq. \eq{imhb} gives $\partial_{M^2}\bar U_2=0$.
Polarization effects do not appear for a homogeneous field.

The non-analyticity of the effective action at $\omega=p=0$ is due to
the photon sector only. The electron sector can be studied in the
homogeneous photon field approximation, $\eta=0$ in Eq.~(\ref{phf}).
The identification of the contributions $\ord{\psid\psi}$ in
Eq.~(\ref{eveqg}), the use of Eqs~(\ref{eq:Aq}) and~(\ref{elevh})
taken at the one-loop approximation and the initial condition
Eq.~(\ref{initc}) lead to
\bea\label{imhf}
\partial_{M^2}[X_0\omega_q-X_1\ke_q+U_1]&\grad&i\hbar
f'e^2\int_p{\ke_p\over(\hbar\omega_p-(1+f)\ke_p+eu_0)^2((p+q)^2+M^2)}\nonu
&&+i\hbar e^2\int_p{1\over \left(\hbar\omega_p-(1+f)\ke_p+eu_0\right)
\left((p+q)^2+M^2\right)^2}.
\eea

The frequency integral in the first term appearing in Eq. \eq{imhf}
vanishes. The second term gives
\be
-e^2\int{d^3p\over(2\pi)^3}{\Theta(-\ke_p)\over
\left((p+q)^2+M^2\right)^2} =-{e^2\over 2\pi^2}\int_0^{k_F}
{p^2dp\over (p^2+q^2+M^2)^2-4p^2q^2}.
\ee
In order to identify the $\ord{\omega_q}$, $\ord{\ke_q}$ and $\ord{1}$
terms with those appearing in the left hand side of Eq.~(\ref{imhf}),
the last expression is expanded in powers of $\ke_q$,
\be
-{e^2\over2\pi^2}\int_0^{k_F}{p^2dp\over
(p^2+k_F^2+M^2)^2-4p^2k_F^2}\left(1-{4m(-p^2+k_F^2+M^2)\over
(p^2+k_F^2+M^2)^2-4p^2k_F^2}\ke_q+\ord{\ke_q^2}\right).
\ee
The identification of the different terms in Eq.~(\ref{imhf}) leads to
the evolution equations $\partial_{M^2}X_0=0$, and
\bea
\partial_{M^2}X_1&=&{-2m e^2\over\pi^2}\int_0^{k_F}{p^2dp(-p^2+k_F^2+M^2)
\over\left((p^2+k_F^2+M^2)^2-4p^2k_F^2\right)^2},\nonu
\partial_{M^2}U_1&=&{-e^2\over 2\pi^2}\int_0^{k_F}
{p^2dp\over(p^2+k_F^2+M^2)^2-4p^2k_F^2}.
\eea
The right hand side of these equations is $u_0$-independent, giving
$X_{0,M^2}=\hbar$, $X_{1,M^2}(u_0)=X_{1,M^2}$ and $U_{1,M^2}(u_0)=U_{1,M^2}=eu_0=0$
with
\be
\partial_{M^2}X_{1,M^2}=-\frac{me^2}{8\pi^2}\left\{
\frac{1}{2k_F^3}\ln\frac{M^2}{4k_F^2+M^2}+\frac{1}{k_F(4k_F^2+M^2)}
+\frac{1}{k_FM^2}\right\}
\ee
In order to obtain the expression for $X_{1,M^2}$, one has to integrate
this diffential equation over $M^2$ from $M^2\to\infty$ to
$\epsilon\to0$.  This yields
\be
X_{1,M^2}=-\frac{me^2}{8\pi^2}\frac{1}{k_F}
\left(2+\lim_{\epsilon\to0}\ln\frac{\epsilon}{4k_F^2}\right)+\ord{M^{-2}}.
\ee
This coupling constant represents the ratio of the bare to the
effective mass of the electron. It diverges as
$\epsilon\to0$ meaning that the electron effective mass vanishes.
This is a well known result from perturbation theory \cite{fetter}
and corresponds to the Hartree-Fock approximation.  Note that the
divergence in the perturbation expansion is also logarithmic and appears as
$\lim_{k\to k_F}\ln(k-k_F)/2k_F$.   The singularity appearing when 
$k\to k_F$ corresponds in the $M$-parameter space to $M\to 0$.

\subsection{Fluctuations: Thomas-Fermi mass}
The reason to go beyond the homogeneous photon field approximation
is the need of the kinetic energy for the photon to reproduce the
desired non-analycity in the photon self energy at $\omega=p=0$.
The terms $\ord{\eta^2}$ in the evolution equation determine
the inverse photon propagator,
\bea
&&\hf\left(\partial_{M^2}Z_0\omega^2_q-\partial_{M^2}Z_1q^2
  -\partial_{M^2}U_2^{(2)}\right)\grad\nonu
&&~~~-i\hbar f'e^2\int{d^3p\over(2\pi)^3}\int{d\omega_p\over2\pi}
{\ke_p\over(\hbar\omega_p-(1+f)\ke_p+eu_0)^2
(\hbar\omega_{p+q}-(1+f)\ke_{p+q}+eu_0)}.
\eea
The frequency integral yields
\be
\frac i\hbar \ke_p
\frac{-\Theta(-\ke_p)+\Theta(-\ke_{p+q})}{\left(\hbar\omega_q-(1+f)
(\ke_{p+q}-\ke_p)+i(\delta_{p+q}-\delta_p)\right)^2}
\ee
which can be written as
\be
\frac i\hbar \ke_p\left\{
\frac{-\Theta(-\ke_p)\Theta(\ke_{p+q})}{\left(\hbar\omega_q-(1+f)
(\ke_{p+q}-\ke_p)+i\delta\right)^2}
+\frac{\Theta(\ke_p)\Theta(-\ke_{p+q})}{\left(\hbar\omega_q-(1+f)
(\ke_{p+q}-\ke_p)-i\delta\right)^2}\right\},
\ee
remembering that $\delta_p=sign(\ke_p)\delta$.
Performing the change of variable $-p-q\to p$ in the second term, one finds
\bea \label{eq:polaris}
&&\hf\Big(\partial_{M^2}Z_0\omega^2_q-\partial_{M^2}Z_1q^2
-\partial_{M^2}U_2^{(2)}\Big)\grad
f'e^2\int{d^3p\over(2\pi)^3}\Theta(-\ke_p)[1-\Theta(-\ke_{p+q})]\nonu
&&~~~\times\left\{\frac{-\ke_p}{\left(
\hbar\omega_q-(1+f)(\ke_{p+q}-\ke_p)+i\delta\right)^2}
+\frac{\ke_{p+q}}{\left(\hbar\omega_q+(1+f)(\ke_{p+q}-\ke_p)
-i\delta\right)^2}\right\}.
\eea
The dependence on $\omega_q$ drops out as $q\to0$, giving $Z_0=\hbar$.

In the following, the real and imaginary parts are treated
separately.  The imaginary part is easier to compute once the
integration over $M^2$ has been performed. We find
\be
\pi e^2\int{d^3p\over(2\pi)^3}\frac{\Theta(-\ke_p)\Theta(\ke_{p+q})}
{\ke_{p+q}-\ke_p}\left\{\ke_p \delta(\omega_q-\ke_{p+q}+\ke_p)
-\ke_{p+q}\delta(\omega_q+\ke_{p+q}-\ke_p)\right\}
\ee
which is vanishing when $\omega_q=0$ is imposed.

The real part of the r.h.s. of Eq.~(\ref{eq:polaris}) is simplified by
noticing that the exchange
$p\rightarrow^{\hskip -.4cm\displaystyle\leftarrow}p+q$
leads to the vanishing of the products of Heaviside step functions.
Moreover, setting $\omega_q=0$ yields
\be\label{lijon}
\partial_{M^2}Z_1q^2+\partial_{M^2}U_2^{(2)}\grad
-2e^2{f'\over (1+f)^2}\int{d^3p\over(2\pi)^3}{\Theta(-\ke_p)\over
(\ke_{p+q}-\ke_p)}
\grad-{f'\over (1+f)^2}{me^2k_F\over2\pi^2}g\left({q\over k_F}\right),
\ee
where
\be
g(x)=\hf-{1\over2x}\left(1-{x^2\over4}\right)\ln\left|
{1-{x\over2}\over1+{x\over2}}\right|
\ee
is the Lindhard function \cite{fetter}.  The integration over $M^2$
from $M\to\infty$ to 0 is easily computed.  The expansion in powers of
$q$ finally gives
\be
Z_{1,M^2}(u_0)=-\frac{me^2}{24\pi^2k_F}+\ord{M^{-2}},~~~
U_{2,M^2}(u_0)=\frac{mk_F e^2}{4\pi^2}u_0^2+\ord{M^{-2}}.
\ee
The non-vanishing coefficient of $u_0^2$, $mk_F e^2/4\pi^2$,
is the Thomas-Fermi mass giving the screening length in the electron
gas. The other coefficient functions remain unchanged at the one-loop
approximation.

\section{Summary}
A new non-perturbative evolution scheme is developed in this paper for
the electron gas with Coulomb interaction. No expansion in the electric
charge is needed, the only approximation
is the restriction of the evolution of the effective action to
a given subspace of functionals, i.e. the consideration
of a given set of 1PI functions with a given energy-momentum
dependence. The evolution equations, a system of coupled equations
for the 1PI functions are to be integrated
out numerically from the initial condition given by the tree-level
action functional.

Since the evolution is performed by adjusting the amplitude of
the quantum fluctuations there is no conflict between
the successive elimination strategy of the renormalization group
and the existence of a non-trivial Fermi surface.

The unique feature of this scheme is that by truncating
the effective action at higher order in the fields or in the
derivatives one obtains systematically improved, non-perturbative
approximations. Since the computation is carried out in continuous
space-time and the time can
be chosen to be real this method offers an alternative for
the usual numerical simulations of field theories on the lattice.

The method provides the solution of the theory in the sense that
physically important quantities are easy to express
in terms of 1PI functions, the final result of the evolution equation.
The only approximation, the truncation of the effective action
in the power of the fields and their gradient has transparent
interpretation and can be controled numerically. In fact, the
maximal power of a given field in the effective action
gives order of the 1PI functions retained in "turning on"
the interactions. The order of the gradient expansion corresponds
to the order of the Taylor expansion of the 1PI functions
around the Fermi surface and $p=\omega=0$ for fermions and bosons,
respectively.

We considered in this work the independent mode solution
of the evolution equation only. This approximation reproduces the
one-loop results of the usual perturbation expansion without relying
Feynman graphs. In order to go beyond well known results one has
to consider the system of coupled differential equations for the
coefficient functions in the effective action and integrate them out
numerically. This work, the numerical determination of the
transport coefficients for the Coulomb gas is in progress.

\acknowledgments
This work is supported in part by the grant OTKA T29927/98.

\begin{appendix}
\section{Functional derivatives}
For the right hand side of the evolution equation we find the
second functional derivative matrix of the effective action,
$\Gamma^{(2)}[\psid,\psi,u]$. Since we expand the
effective action in the fluctuations in order to identify the
terms $\ord{\partial^2}$ we ultimately need the functional derivatives
up to fourth order.

The first four derivatives of the effective action \eq{grexf}
written as $\Gamma=\int_x(\psid_xK_x\psi_x+L_x)$,
are
\bea\label{firstfd}
\Gamma^{(1)}_{\psi_y}&=&-\int_x\psid_xK_x\delta_{x,y},\nonu
\Gamma^{(1)}_{\psid_y}&=&\int_x\delta_{x,y}K_x\psi_x,\nonu
\Gamma^{(1)}_{u_y}&=&\int_x\delta_{x,y}(\psid_xK_x^{(1)}\psi_x
+L_x^{(1)})+\int_x\biggl\{\psid_x\biggl[Y_0\partial_tu_x\partial_t\delta_{x,y}
+Y_1\nabla u_x\nabla\delta_{x,y}\biggr]\psi_x\nonu
&&+Z_0\partial_tu_x\partial_t\delta_{x,y}-Z_1\nabla u_x\nabla\delta_{x,y}\biggr\},
\eea
where
\bea
K_x^{(n)}&=&iX_0^{(n)}(u_x)\partial_t-X_1^{(n)}(u_x)\ke+U_1^{(n)}(u_x)+
\hf Y_0^{(n)}(u_x)(\partial_tu_x)^2+\hf Y_1^{(n)}(u_x)(\nabla u_x)^2,\nonu
L^{(n)}_x&=&\hf Z_0^{(n)}(u_x)(\partial_tu_x)^2
-\hf Z^{(n)}_1(u_x)(\nabla u_x)^2-U_2^{(n)}(u_x)
\eea
and the coefficient functions $X$, $Y$ and $Z$ are considered
at the value $u_x$ in the integrands.

The second derivatives,
\bea\label{secondfd}
\Gamma^{(2)}_{\psid_{y_1}\psi_{y_2}}&=&-\int_x\delta_{x,y_1}K_x\delta_{x,y_2},\nonu
\Gamma^{(2)}_{\psi_{y_1}u_{y_2}}&=&-\int_x\delta_{x,y_2}\psid_xK_x^{(1)}\delta_{x,y_1}
-\int_x\psid_x\biggl[Y_0\partial_tu_x\partial_t\delta_{x,y_2}
+Y_1\nabla u_x\nabla\delta_{x,y_2}\biggr]\delta_{x,y_1},\nonu
\Gamma^{(2)}_{\psid_{y_1}u_{y_2}}&=&\int_x\delta_{x,y_1}\delta_{x,y_2}
K_x^{(1)}\psi_x+\int_x\delta_{x,y_1}
\biggl[Y_0\partial_tu_x\partial_t\delta_{x,y_2}
+Y_1\nabla u_x\nabla\delta_{x,y_2}\biggr]\psi_x,\nonu
\Gamma^{(2)}_{u_{y_1}u_{y_2}}&=&\int_x\delta_{x,y_1}\delta_{x,y_2}
(\psid_xK_x^{(2)}\psi_x+L^{(2)}_x)\nonu
&&+\int_x\left(\psid_x\psi_xY_0^{(1)}+Z_0^{(1)}\right)
\partial_tu_x\left(\delta_{x,y_1}\partial_t\delta_{x,y_2}
+\delta_{x,y_2}\partial_t\delta_{x,y_1}\right)\nonu
&&+\int_x\left(\psid_x\psi_xY_1^{(1)}-Z_1^{(1)}\right)
\nabla u_x\left(\delta_{x,y_1}\nabla\delta_{x,y_2}
+\delta_{x,y_2}\nabla_t\delta_{x,y_1}\right)\nonu
&&+\int_x\left(\psid_x\psi_xY_0+Z_0\right)
\partial_t\delta_{x,y_1}\partial_t\delta_{x,y_2}
+\int_x\left(\psid_x\psi_xY_1-Z_1\right)
\nabla\delta_{x,y_1}\nabla\delta_{x,y_2},
\eea
yield the non-diagonal, fluctuation-independent
contribution to the self energy,
\be
\Gamma_{\tn}=\pmatrix{0&0&\Sigma^{\psid u}\cr0&0&\Sigma^{\psi u}\cr
\Sigma^{u\psi}&\Sigma^{u\psid}&\Sigma^{uu}}
\ee
with
\bea\label{abcde}
\Sigma^{u,\psi}_{p_1,p_2}&=&\Sigma^{\psi,u}_{p_2,p_1}
=-\psid_{-p_2-p_1}A_{p_2}^{(1)},\nonu
\Sigma^{u,\psid}_{p_1,p_2}&=&\Sigma^{\psid,u}_{p_2,p_1}
=A_{-p_1-p_2}^{(1)}\psi_{-p_1-p_2},\nonu
\Sigma^{u,u}_{p_1,p_2}&=&\int_qB_{p_1,p_2}^q\psid_{-q-p_1-p_2}\psi_q,
\eea
and
\be
B_{p_1,p_2}^q=A_q^{(2)}-Y_0\omega_{p_1}\omega_{p_2}-Y_1p_1\cdot p_2,
\ee
and $A$ being defined in Eq. \eq{eq:Aq}.
The coefficient functions $X$, $Y$ and $Z$ are considered at $u_0$
in the gradient expansion formulae.

In order to find the contributions of the fluctuations to the self
energy up to the quadratic order we need the following functional
derivatives,
\bea
\Gamma^{(3)}_{\psid_{y_1}\psi_{y_2}u_{y_3}}
&=&-\int_x\delta_{x,y_1}\delta_{x,y_3}K_x^{(1)}\delta_{x,y_2}
-\int_x\delta_{x,y_1}\biggl[Y_0\partial_tu_x\partial_t\delta_{x,y_3}
+Y_1\nabla u_x\nabla\delta_{x,y_3}\biggr]\delta_{x,y_2},\nonu
\Gamma^{(3)}_{\psi_{y_1}u_{y_2}u_{y_3}}
&=&-\int_x\delta_{x,y_2}\delta_{x,y_3}\psid_xK_x^{(2)}\delta_{x,y_1}
-\int_x\psid_xY^{(1)}_0\partial_tu_x\delta_{x,y_1}\left(
\delta_{x,y_3}\partial_t\delta_{x,y_2}
+\delta_{x,y_2}\partial_t\delta_{x,y_3}\right)\nonu
&&-\int_x\psid_xY^{(1)}_1\nabla u_x\delta_{x,y_1}\left(
\delta_{x,y_3}\nabla\delta_{x,y_2}
+\delta_{x,y_2}\nabla\delta_{x,y_3}\right)\nonu
&&-\int_x\psid_x\biggl[Y_0\partial_t\delta_{x,y_3}
\partial_t\delta_{x,y_2}+Y_1\nabla\delta_{x,y_3}
\nabla\delta_{x,y_2}\biggr]\delta_{x,y_1},\nonu
\Gamma^{(3)}_{\psid_{y_1}u_{y_2}u_{y_3}}
&=&\int_x\delta_{x,y_1}\delta_{x,y_2}\delta_{x,y_3}K_x^{(2)}\psi_x
+\int_x\delta_{x,y_1}Y^{(1)}_0\partial_tu_x\left(
\delta_{x,y_3}\partial_t\delta_{x,y_2}
+\delta_{x,y_2}\partial_t\delta_{x,y_3}\right)\psi_x\nonu
&&+\int_x\delta_{x,y_1}Y^{(1)}_1\nabla u_x\left(
\delta_{x,y_3}\nabla\delta_{x,y_2}
+\delta_{x,y_2}\nabla\delta_{x,y_3}\right)\psi_x\nonu
&&+\int_x\delta_{x,y_1}\biggl[Y_0\partial_t\delta_{x,y_3}
\partial_t\delta_{x,y_2}+Y_1\nabla\delta_{x,y_3}
\nabla\delta_{x,y_2}\biggr]\psi_x,\nonu
\Gamma^{(3)}_{u_{y_1}u_{y_2}u_{y_3}}
&=&\int_x\delta_{x,y_1}\delta_{x,y_2}\delta_{x,y_3}(\psid_xK_x^{(3)}\psi_x+L^{(3)}_x)
+\int_x\left(\psid_x\psi_xY^{(2)}_0+Z^{(2)}_0\right)\partial_tu_x\nonu
&&\times\left(\delta_{x,y_2}\delta_{x,y_3}\partial_t\delta_{x,y_1}
+\delta_{x,y_1}\delta_{x,y_3}\partial_t\delta_{x,y_2}
+\delta_{x,y_1}\delta_{x,y_2}\partial_t\delta_{x,y_3}\right)\nonu
&&+\int_x\left(\psid_x\psi_xY^{(2)}_1-Z^{(2)}_1\right)\nabla u_x\nonu
&&\times\left(\delta_{x,y_2}\delta_{x,y_3}\nabla\delta_{x,y_1}
+\delta_{x,y_1}\delta_{x,y_3}\nabla\delta_{x,y_2}
+\delta_{x,y_1}\delta_{x,y_2}\nabla\delta_{x,y_3}\right)\nonu
&&+\int_x\left(\psid_x\psi_xY^{(1)}_0+Z^{(1)}_0\right)\nonu
&&\times\left(\delta_{x,y_1}\partial_t\delta_{x,y_2}\partial_t\delta_{x,y_3}
+\delta_{x,y_2}\partial_t\delta_{x,y_3}\partial_t\delta_{x,y_1}
+\delta_{x,y_3}\partial_t\delta_{x,y_1}\partial_t\delta_{x,y_2}\right)\nonu
&&+\int_x\left(\psid_x\psi_xY_1^{(1)}-Z_1^{(1)}\right)\nonu
&&\times\left(\delta_{x,y_1}\nabla\delta_{x,y_2}\nabla\delta_{x,y_3}
+\delta_{x,y_2}\nabla\delta_{x,y_3}\nabla\delta_{x,y_1}
+\delta_{x,y_3}\nabla\delta_{x,y_1}\nabla\delta_{x,y_2}\right),
\eea

\bea
\Gamma^{(4)}_{\psid_{y_1}\psi_{y_2}u_{y_3}u_{y_4}}
&=&-\int_x\delta_{x,y_1}\delta_{x,y_3}\delta_{x,y_4}K_x^{(2)}\delta_{x,y_2}
-\int_x\delta_{x,y_1}\delta_{x,y_2}Y^{(1)}_0\partial_tu_x\left(
\delta_{x,y_3}\partial_t\delta_{x,y_4}
+\delta_{x,y_4}\partial_t\delta_{x,y_3}\right)\nonu
&&-\int_x\delta_{x,y_1}\delta_{x,y_2}Y^{(1)}_1  \nabla u_x\left(
\delta_{x,y_3}\nabla\delta_{x,y_4}
+\delta_{x,y_4}\nabla\delta_{x,y_3}\right)\nonu
&&-\int_x\delta_{x,y_1}\delta_{x,y_2}\left(
Y_0\partial_t\delta_{x,y_3}\partial_t\delta_{x,y_4}
+Y_1\nabla\delta_{x,y_3}\nabla\delta_{x,y_4}\right),\nonu
\Gamma^{(4)}_{u_{y_1}u_{y_2}u_{y_3}\psid_{y_4}}
&=&\int_x\delta_{x,y_1}\delta_{x,y_2}\delta_{x,y_3}\delta_{x,y_4}
K_x^{(3)}\psi_x\nonu
&&+\int_x\psi_x\delta_{x,y_4}Y^{(2)}_0\partial_tu_x
\left(\delta_{x,y_2}\delta_{x,y_3}\partial_t\delta_{x,y_1}
+\delta_{x,y_1}\delta_{x,y_3}\partial_t\delta_{x,y_2}
+\delta_{x,y_1}\delta_{x,y_2}\partial_t\delta_{x,y_3}\right)\nonu
&&+\int_x\psi_x\delta_{x,y_4}Y^{(2)}_1\nabla u_x
\left(\delta_{x,y_2}\delta_{x,y_3}\nabla\delta_{x,y_1}
+\delta_{x,y_1}\delta_{x,y_3}\nabla\delta_{x,y_2}
+\delta_{x,y_1}\delta_{x,y_2}\nabla\delta_{x,y_3}\right)\nonu
&&+\int_x\psi_x\delta_{x,y_4}Y^{(1)}_0
\left(\delta_{x,y_1}\partial_t\delta_{x,y_2}\partial_t\delta_{x,y_3}
+\delta_{x,y_2}\partial_t\delta_{x,y_3}\partial_t\delta_{x,y_1}
+\delta_{x,y_3}\partial_t\delta_{x,y_1}\partial_t\delta_{x,y_2}\right)\nonu
&&+\int_x\psi_x\delta_{x,y_4}Y_1^{(1)}
\left(\delta_{x,y_1}\nabla\delta_{x,y_2}\nabla\delta_{x,y_3}
+\delta_{x,y_2}\nabla\delta_{x,y_3}\nabla\delta_{x,y_1}
+\delta_{x,y_3}\nabla\delta_{x,y_1}\nabla\delta_{x,y_2}\right),\nonu
\Gamma^{(4)}_{u_{y_1}u_{y_2}u_{y_3}\psi_{y_4}}
&=&-\int_x\delta_{x,y_1}\delta_{x,y_2}\delta_{x,y_3}
\psid_xK_x^{(3)}\delta_{x,y_4}\nonu
&&-\int_x\psid_x\delta_{x,y_4}Y^{(2)}_0\partial_tu_x
\left(\delta_{x,y_2}\delta_{x,y_3}\partial_t\delta_{x,y_1}
+\delta_{x,y_1}\delta_{x,y_3}\partial_t\delta_{x,y_2}
+\delta_{x,y_1}\delta_{x,y_2}\partial_t\delta_{x,y_3}\right)\nonu
&&-\int_x\psid_x\delta_{x,y_4}Y^{(2)}_1\nabla u_x
\left(\delta_{x,y_2}\delta_{x,y_3}\nabla\delta_{x,y_1}
+\delta_{x,y_1}\delta_{x,y_3}\nabla\delta_{x,y_2}
+\delta_{x,y_1}\delta_{x,y_2}\nabla\delta_{x,y_3}\right)\nonu
&&-\int_x\psid_x\delta_{x,y_4}Y^{(1)}_0
\left(\delta_{x,y_1}\partial_t\delta_{x,y_2}\partial_t\delta_{x,y_3}
+\delta_{x,y_2}\partial_t\delta_{x,y_3}\partial_t\delta_{x,y_1}
+\delta_{x,y_3}\partial_t\delta_{x,y_1}\partial_t\delta_{x,y_2}\right)\nonu
&&-\int_x\psid_x\delta_{x,y_4}Y_1^{(1)}
\left(\delta_{x,y_1}\nabla\delta_{x,y_2}\nabla\delta_{x,y_3}
+\delta_{x,y_2}\nabla\delta_{x,y_3}\nabla\delta_{x,y_1}
+\delta_{x,y_3}\nabla\delta_{x,y_1}\nabla\delta_{x,y_2}\right),\nonu
\Gamma^{(4)}_{u_{y_1}u_{y_2}u_{y_3}u_{y_4}}
&=&\int_x\delta_{x,y_1}\delta_{x,y_2}\delta_{x,y_3}\delta_{x,y_4}
(\psid_xK_x^{(4)}\psi_x+L^{(4)}_x)\nonu
&&+\int_x\left(\psid_x\psi_xY^{(3)}_0+Z^{(3)}_0\right)\partial_tu_x
\biggl(\delta_{x,y_2}\delta_{x,y_3}\delta_{x,y_4}\partial_t\delta_{x,y_1}\nonu
&&+\delta_{x,y_1}\delta_{x,y_3}\delta_{x,y_4}\partial_t\delta_{x,y_2}
+\delta_{x,y_1}\delta_{x,y_2}\delta_{x,y_4}\partial_t\delta_{x,y_3}
+\delta_{x,y_1}\delta_{x,y_2}\delta_{x,y_3}\partial_t\delta_{x,y_4}\biggr)\nonu
&&+\int_x\left(\psid_x\psi_xY^{(3)}_1-Z^{(3)}_1\right)\nabla u_x
\biggl(\delta_{x,y_2}\delta_{x,y_3}\delta_{x,y_4}\nabla\delta_{x,y_1}\nonu
&&+\delta_{x,y_1}\delta_{x,y_3}\delta_{x,y_4}\nabla\delta_{x,y_2}
+\delta_{x,y_1}\delta_{x,y_2}\delta_{x,y_4}\nabla\delta_{x,y_3}
+\delta_{x,y_1}\delta_{x,y_2}\delta_{x,y_3}\nabla\delta_{x,y_4}\biggr)\nonu
&&+\int_x\left(\psid_x\psi_xY^{(2)}_0+Z^{(2)}_0\right)
\biggl(\delta_{x,y_1}\delta_{x,y_4}\partial_t\delta_{x,y_2}\partial_t\delta_{x,y_3}\nonu
&&+\delta_{x,y_2}\delta_{x,y_4}\partial_t\delta_{x,y_3}\partial_t\delta_{x,y_1}
+\delta_{x,y_2}\delta_{x,y_3}\partial_t\delta_{x,y_1}\partial_t\delta_{x,y_4}
+\delta_{x,y_1}\delta_{x,y_3}\partial_t\delta_{x,y_2}\partial_t\delta_{x,y_4}\nonu
&&+\delta_{x,y_1}\delta_{x,y_2}\partial_t\delta_{x,y_3}\partial_t\delta_{x,y_4}
+\delta_{x,y_3}\delta_{x,y_4}\partial_t\delta_{x,y_1}\partial_t\delta_{x,y_2}\biggr)\nonu
&&+\int_x\left(\psid_x\psi_xY_1^{(2)}-Z_1^{(2)}\right)
\biggl(\delta_{x,y_1}\delta_{x,y_4}\nabla\delta_{x,y_2}\nabla\delta_{x,y_3}
+\delta_{x,y_2}\delta_{x,y_4}\nabla\delta_{x,y_3}\nabla\delta_{x,y_1}\nonu
&&+\delta_{x,y_2}\delta_{x,y_3}\nabla\delta_{x,y_1}\nabla\delta_{x,y_4}
+\delta_{x,y_1}\delta_{x,y_3}\nabla\delta_{x,y_2}\nabla\delta_{x,y_4}\nonu
&&+\delta_{x,y_1}\delta_{x,y_2}\nabla\delta_{x,y_3}\nabla\delta_{x,y_4}
+\delta_{x,y_3}\delta_{x,y_4}\nabla\delta_{x,y_1}\nabla\delta_{x,y_2}\biggr).
\eea

The matrix elements \eq{parsecse} needed for the right hand side of the
evolution equation can be read off easily,
\bea\label{fluctsee}
\sigma^{\psid\psi}_{p_1,p_2}&=&-\sigma^{\psi\psid}_{p_2,p_1}
=-A_{p_2}^{(1)}\eta_{-p_1-p_2}-\int_qC_{p_1,p_2}^q
\eta_{-q-p_1-p_2}\eta_q,\nonu
\sigma^{\psid\psid}_{p_1,p_2}&=&\sigma^{\psi\psi}_{p_1,p_2}=0,\nonu
\sigma^{\psid u}_{p_1,p_2}&=&\sigma^{u\psid}_{p_2,p_1}
=\int_qD_{p_1,p_2}^q\eta_q\psi_{-q-p_1-p_2}
+\int_{q_1,q_2}E_{p_1,p_2}^{q_1,q_2}
\eta_{q_1}\eta_{q_2}\psi_{-q_1-q_2-p_1-p_2},\nonu
\sigma^{\psi u}_{p_1,p_2}&=&\sigma^{u\psi}_{p_2,p_1}
=-\int_qF_{p_1,p_2}^q\eta_q\psid_{-q-p_1-p_2}
-\int_{q_1,q_2}G_{p_1,p_2}^{q_1,q_2}\eta_{q_1}\eta_{q_2}\psid_{-q_1-q_2-p_1-p_2},\nonu
\sigma^{uu}_{p_1,p_2}&=&-H_{p_1,p_2}\eta_{-p_1-p_2}
+\int_{q_1,q_2}I_{p_1,p_2}^{q_1,q_2}\eta_{q_1}\psid_{q_2}\psi_{-q_1-q_2-p_1-p_2}\nonu
&&+\int_{q_1,q_2,q_3}J_{p_1,p_2}^{q_1,q_2,q_3}\eta_{q_1}\eta_{q_2}\psid_{q_3}
\psi_{-q_1-q_2-q_3-p_1-p_2}-\int_qK_{p_1,p_2}^q\eta_q\eta_{-q-p_1-p_2},
\eea
where
\bea
C_{p_1,p_2}^q&=&\hf\left[A_{p_2}^{(2)}-Y_0\omega_q(\omega_q+\omega_{p_1}
+\omega_{p_2})+Y_1q\cdot(q+p_1+p_2)\right],\nonu
D_{p_1,p_2}^q&=&A_{-q-p_1-p_2}^{(2)}-Y_0\omega_q\omega_{p_2}-Y_1q\cdot p_2,\nonu
E_{p_1,p_2}^{q_1,q_2}&=&\hf\left\{A_{-q_1-q_2-p_1-p_2}^{(3)}
-Y_0^{(1)}\left[\omega_{q_1}\omega_{q_2}+(\omega_{q_1}+\omega_{q_2})
\omega_{p_2}\right]-Y_1^{(1)}\left[q_1\cdot q_2+(q_1+q_2)\cdot p_2\right]\right\},\nonu
F_{p_1,p_2}^q&=&A_{p_1}^{(2)}-Y_0\omega_q\omega_{p_2}-Y_1q\cdot p_2,\nonu
G_{p_1,p_2}^{q_1,1q_2}&=&\hf\left\{A_{p_1}^{(3)}
-Y_0^{(1)}\left[\omega_{p_2}(\omega_{q_1}+\omega_{q_2})
+\omega_{q_1}\omega_{q_2}\right]-Y_1^{(1)}\left[p_2\cdot(q_1+q_2)
+q_1\cdot q_2\right]\right\},\nonu
H_{p_1,p_2}&=&Z_0^{(1)}\left[
(\omega_{p_1}+\omega_{p_2})^2+\omega_{p_1}\omega_{p_2}\right]
+Z_1^{(1)}\left[(p_1+p_2)^2+p_1\cdot p_2\right]+U^{(3)}_2,\nonu
I_{p_1,p_2}^{q_1,q_2}&=&A_{-q_1-q_2-p_1-p_2}^{(3)}-Y_0^{(1)}
\left[\omega_{q_1}(\omega_{p_1}+\omega_{p_2})+\omega_{p_1}\omega_{p_2}\right]
-Y_1^{(1)}\left[q_1\cdot(p_1+p_2)+p_1\cdot p_2\right],\nonu
J_{p_1,p_2}^{q_1,q_2,q_3}&=&\hf\biggl\{A_{-q_1-q_2-q_3-p_1-p_2}^{(4)}-Y_0^{(2)}
\left[\omega_{q_1}\omega_{q_2}+(\omega_{q_1}+\omega_{q_2})
(\omega_{p_1}+\omega_{p_2})+\omega_{p_1}\omega_{p_2}\right]\nonu
&&-Y_1^{(2)}\left[q_1\cdot q_2+(q_1+q_2)\cdot(p_1+p_2)
+p_1\cdot p_2\right]\biggr\},\nonu
K_{p_1,p_2}^q&=&\hf\biggl\{Z_0^{(2)}\left[(\omega_q+\omega_{p_1}+\omega_{p_2})^2
+\omega_q(\omega_{p_1}+\omega_{p_2})+\omega_{p_1}\omega_{p_2}\right]\nonu
&&+Z_1^{(2)}\left[(q+p_1+p_2)^2-q\cdot(p_1+p_2)-p_1\cdot p_2\right]
+U_2^{(4)}\biggr\}.
\eea

\section{Gradient expansion}\label{neumap}
Here we provide more detailed expressions for the gradient expansion of
the evolution equation Eq. \eq{eveqg}.

The fluctuations of the photon field to the evolution equation
will be retaind up to the level $\ord{\psid\psi\eta^2}$.
These contributions from the expansion \eq{neum}
are inserted on the right hand side of the evolution equation
Eq. \eq{eveqg}. We introduce the notation
\be
{\cal F}_n=-\sum_{\alpha,\cdots,\alpha_n}
\tr\left[\ke W_{\alpha_1,\cdots,\alpha_n,j,\jd}\right],~~~
{\cal B}_n=\sum_{\alpha,\cdots,\alpha_n}
\tr\left[W_{\alpha_1,\cdots,\alpha_n,J,J}\right],
\ee
where the summation is over index sets where the number of value
"fl" is 1 or 2. We find for the necessary terms
\bea\label{neumel}
-\tr\left[\ke W_{\tf,j,\jd}\right] 
&=&\int_p\ke_pG^2_p\sigma^{\psi\psid}_{-p,p},\nonu
{\cal F}_2&=&-\int_{p,p'}\ke_pG^2_p(
\Sigma^{\psi u}_{-p,p'}D_{p'}\sigma^{u\psid}_{-p',p}
+\sigma^{\psi u}_{-p,p'}D_{p'}\Sigma^{u\psid}_{-p',p}
+\sigma^{\psi\psid}_{-p,p'}G_{-p'}\sigma^{\psi\psid}_{-p',p}
+\sigma^{\psi u}_{-p,p'}D_{p'}\sigma^{u\psid}_{-p',p}),\nonu
{\cal F}_3&=&\int_{p,p',p''}\ke_pG^2_p(
\sigma^{\psi\psid}_{-p,p'}G_{-p'}\Sigma^{\psi u}_{-p',p''}
D_{p''}\Sigma^{u\psid}_{-p'',p}+\Sigma^{\psi u}_{-p,p'}
D_{p'}\sigma^{uu}_{-p',p''}D_{p''}\Sigma^{u\psid}_{-p'',p}\nonu
&&+\Sigma^{\psi u}_{-p,p'}D_{p'}\Sigma^{u\psid}_{-p',p''}
G_{-p''}\sigma^{\psi\psid}_{-p'',p}
+\sigma^{\psi\psid}_{-p,p'}G_{-p'}\sigma^{\psi u}_{-p',p''}
D_{p''}\Sigma^{u\psid}_{-p'',p}
+\sigma^{\psi u}_{-p,p'}D_{p'}\sigma^{uu}_{-p',p''}
D_{p''}\Sigma^{u\psid}_{-p'',p}\nonu
&&+\sigma^{\psi\psid}_{-p,p'}G_{-p'}\Sigma^{\psi u}_{-p',p''}
D_{p''}\sigma^{u\psid}_{-p'',p}
+\Sigma^{\psi u}_{-p,p'}D_{p'}
\sigma^{u\psid}_{-p',p''}G_{-p''}\sigma^{\psi\psid}_{-p'',p}
+\Sigma^{\psi u}_{-p,p'}D_{p'}\sigma^{uu}_{-p',p''}D_{p''}
\sigma^{u\psid}_{-p'',p}),\nonu
{\cal F}_4&=&\int_{p,p',p'',p'''}\ke_pG^2_p(
-\sigma^{\psi\psid}_{-p,p'}G_{-p'}\sigma^{\psi\psid}_{-p',p''}
G_{-p''}\Sigma^{\psi u}_{-p'',p'''}D_{p'''}\Sigma^{u\psid}_{-p''',p}\nonu
&&-\sigma^{\psi\psid}_{-p,p'}G_{-p'}\Sigma^{\psi u}_{-p',p''}D_{p''}
\sigma^{uu}_{-p'',p'''}D_{p'''}\Sigma^{u\psid}_{-p''',p}
-\Sigma^{\psi u}_{-p,p'}D_{p'}\sigma^{uu}_{-p',p''}D_{p''}
\sigma^{uu}_{-p'',p'''}D_{p'''}\Sigma^{u\psid}_{-p''',p}\nonu
&&+\Sigma^{\psi u}_{-p,p'}D_{p'}\sigma^{uu}_{-p',p''}D_{p''}
\Sigma^{u\psid}_{-p'',p'''}G_{p'''}\sigma^{\psi\psid}_{-p''',p}
-\Sigma^{\psi u}_{-p,p'}D_{p'}\Sigma^{u\psid}_{-p',p''}G_{-p''}
\sigma^{\psi\psid}_{-p'',p'''}G_{-p'''}\sigma^{\psi\psid}_{-p''',p}),
\eea
for the electron contributions and
\bea\label{neumph}
\tr\left[W_{\tf,J,J}\right]&=&-\int_pD^2_p\sigma^{uu}_{-p,p},\nonu
{\cal B}_2&=&\int_{p,p'}D^2_p(
-2\Sigma^{u\psi}_{-p,p'}G_{p'}\sigma^{\psid u}_{-p',p}
+2\Sigma^{u\psid}_{-p,p'}G_{-p'}\sigma^{\psi u}_{-p',p}
+2\Sigma^{uu}_{-p,p'}D_{p'}\sigma^{uu}_{-p',p}\nonu
&&-\sigma^{u\psi}_{-p,p'}G_{p'}\sigma^{\psid u}_{-p',p}
+\sigma^{u\psid}_{-p,p'}G_{-p'}\sigma^{\psi u}_{-p',p}
+\sigma^{uu}_{-p,p'}D_{p'}\sigma^{uu}_{-p',p}),\nonu
{\cal B}_3&=&\int_{p,p',p''}D^2_p(
\sigma^{uu}_{-p,p'}D_{p'}\Sigma^{u\psi}_{-p',p''}G_{p''}\Sigma^{\psid u}_{-p'',p}
-\sigma^{uu}_{-p,p'}D_{p'}\Sigma^{u\psid}_{-p',p''}G_{-p''}\Sigma^{\psi u}_{-p'',p}
+\Sigma^{u\psi}_{-p,p'}G_{p'}\Sigma^{\psid u}_{-p',p}D_{p''}\sigma^{uu}_{-p'',p}\nonu
&&-\Sigma^{u\psid}_{-p,p'}G_{-p'}\Sigma^{\psi u}_{-p',p}D_{p''}\sigma^{uu}_{-p'',p}
+\Sigma^{u\psi}_{-p,p'}G_{p'}\sigma^{\psid u}_{-p',p''}D_{p''}\sigma^{uu}_{-p'',p}
-\Sigma^{u\psid}_{-p,p'}G_{-p'}\sigma^{\psi u}_{-p',p''}D_{p''}\sigma^{uu}_{-p'',p}\nonu
&&+\sigma^{uu}_{-p,p'}D_{p'}\Sigma^{u\psi}_{-p',p''}G_{p''}\sigma^{\psid u}_{-p'',p}
-\sigma^{uu}_{-p,p'}D_{p'}\Sigma^{u\psid}_{-p',p''}G_{-p''}\sigma^{\psi u}_{-p'',p}
+\sigma^{u\psi}_{-p,p'}G_{p'}\Sigma^{\psid u}_{-p',p''}D_{p''}\sigma^{uu}_{-p'',p}\nonu
&&-\sigma^{u\psid}_{-p,p'}G_{-p'}\Sigma^{\psi u}_{-p',p''}D_{p''}\sigma^{uu}_{-p'',p}
+\sigma^{uu}_{-p,p'}D_{p'}\sigma^{u\psi}_{-p',p''}G_{p''}\Sigma^{\psid u}_{-p'',p}
-\sigma^{uu}_{-p,p'}D_{p'}\sigma^{u\psid}_{-p',p''}G_{-p''}\Sigma^{\psi u}_{-p'',p}),\nonu
{\cal B}_4&=&\int_{p,p',p'',p'''}D^2_p(
2\sigma^{uu}_{-p,p'}D_{p'}\sigma^{uu}_{-p',p''}
D_{p''}\Sigma^{u\psid}_{-p'',p'''}G_{-p'''}\Sigma^{\psi u}_{-p''',p}\nonu
&&-2\sigma^{uu}_{-p,p'}D_{p'}\sigma^{uu}_{-p',p''}
D_{p''}\Sigma^{u\psi}_{-p'',p'''}G_{p'''}\Sigma^{\psid u}_{-p''',p}\nonu
&&+2\sigma^{uu}_{-p,p'}D_{p'}\Sigma^{u\psid}_{-p',p''}
G_{-p''}\sigma^{\psi\psid}_{-p'',p'''}G_{-p'''}\Sigma^{\psi u}_{-p''',p}
+2\sigma^{uu}_{-p,p'}D_{p'}\Sigma^{u\psi}_{-p',p''}
G_{p''}\sigma^{\psid\psi}_{-p'',p'''}G_{p'''}\Sigma^{\psid u}_{-p''',p}\nonu
&&+\sigma^{uu}_{-p,p'}D_{p'}\Sigma^{u\psid}_{-p',p''}
G_{-p''}\Sigma^{\psi u}_{-p'',p'''}D_{p'''}\sigma^{uu}_{-p''',p}
-\sigma^{uu}_{-p,p'}D_{p'}\Sigma^{u\psi}_{-p',p''}
G_{p''}\Sigma^{\psid u}_{-p'',p'''}D_{p'''}\sigma^{uu}_{-p''',p}\nonu
&&+\Sigma^{u\psid}_{-p,p'}G_{-p'}\sigma^{\psi\psid}_{-p',p''}
G_{-p''}\sigma^{\psi\psid}_{-p'',p'''}G_{-p'''}\Sigma^{\psi u}_{-p''',p}
-\Sigma^{u\psi}_{-p,p'}G_{p'}\sigma^{\psid\psi}_{-p',p''}
G_{p''}\sigma^{\psid\psi}_{-p'',p'''}G_{p'''}\Sigma^{\psid u}_{-p''',p}),
\eea
for the photon induced terms. As discussed below Eq. \eq{etat}
the sum of the terms listed in Eqs. \eq{neumel} and \eq{neumph}
yields the evolution for $Y_0$ and $Y_1$.

\end{appendix}
\end{document}